\documentclass[epj]{webofc}
\usepackage[varg]{txfonts}
\usepackage{hyperref}

\wocname{EPJ Web of Conferences}

\woctitle{CONF12}

\begin{document}

\selectlanguage{english}
\title{An exploratory study of Yang-Mills three-point functions at non-zero temperature}

\author{Markus Q. Huber\inst{1}\fnsep\thanks{\email{markus.huber@uni-graz.at}}}

\institute{Institute of Physics, University of Graz, NAWI Graz, Universitätsplatz 5, A-8010 Graz, Austria}

\abstract{
Results for three-point functions of Landau gauge Yang-Mills theory at non-vanishing temperature are presented and compared to lattice results. It is found that the three-gluon vertex is enhanced for temperatures below the phase transition. At very low momenta it becomes negative for all temperatures. Furthermore, truncation effects in functional equations are discussed at the example of three-dimensional Yang-Mills theory for which a self-contained solution is presented.
}

\maketitle

\section{Introduction}
\label{sec:introduction}

Despite of decades of dedicated research, the phase diagram of quantum chromodynamics (QCD) is still not fully explored. Especially the phase structure at non-vanishing chemical potential defies a detailed study. The otherwise very successful method of Monte Carlo lattice simulations is severely hampered by the complex action problem appearing for non-vanishing chemical potential, e.g., \cite{deForcrand:2010ys}.
An alternative approach is provided by functional methods. Since this approach is in general less investigated than the lattice approach, some fundamental aspects still need to be clarified. The most pressing one is surely the question of how reliable this method can be on a quantitative level. Recent results indicate that, at least in the vacuum, a reasonable accuracy can be achieved \cite{Mitter:2014wpa,Cyrol:2016tym,Huber:2016tvc}. However, to realize corresponding truncations requires to solve large sets of functional equations.

In this contribution I will report on an analysis of truncation effects on such systems of equations in the vacuum and on results at non-vanishing temperature. The latter are obtained from rather small truncations with the goal to extend them in the future to a similarly sophisticated level as in the vacuum and to generalize them to non-zero chemical potential.

\section{Testing truncations: Yang-Mills theory in three dimensions}
\label{sec:truncations}

\begin{figure}[tb]
 \begin{center}
  \centering \includegraphics[width=\textwidth]{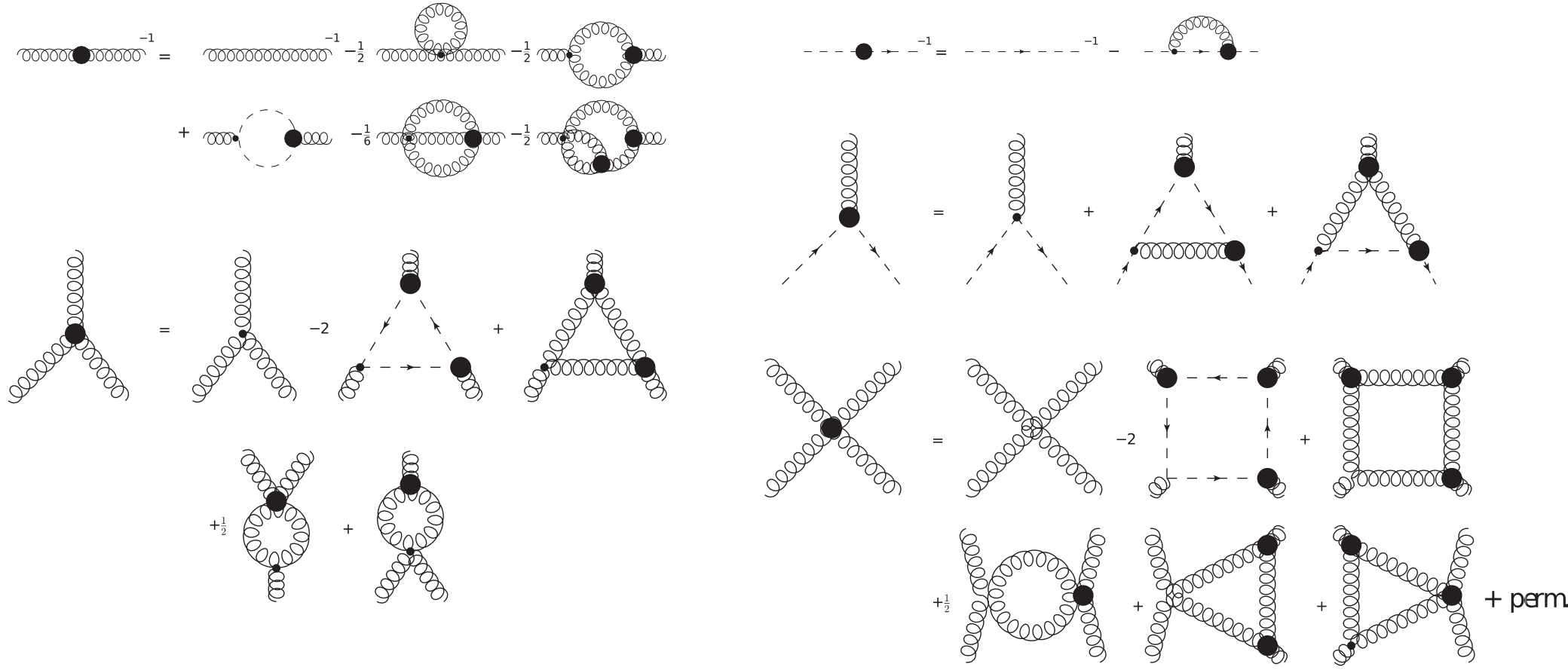}
  \caption{The employed truncated system of equations. Internal propagators are dressed, and thick blobs denote dressed vertices, wiggly lines gluons, and dashed ones ghosts.}
  \label{fig:soe}
 \end{center}
\end{figure}
  
\begin{figure}[b]
 \includegraphics[width=0.49\textwidth]{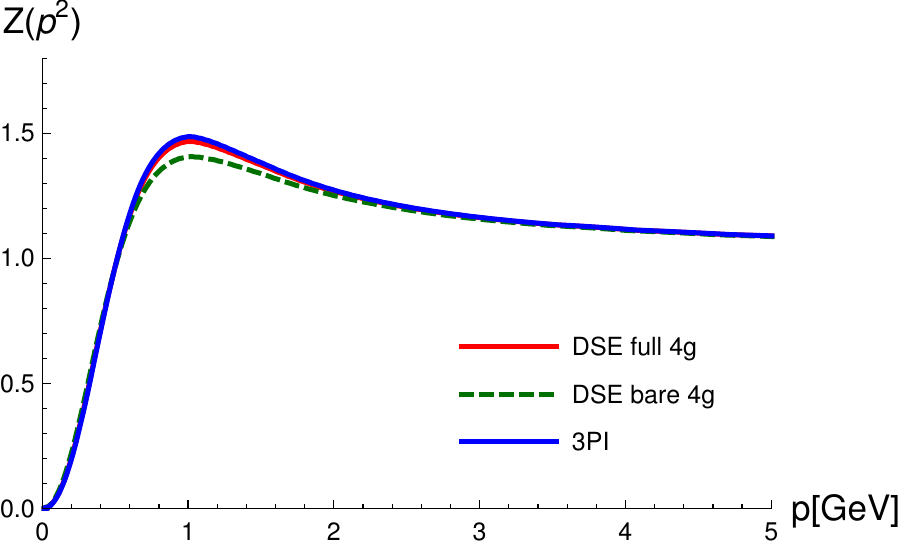}
 \includegraphics[width=0.49\textwidth]{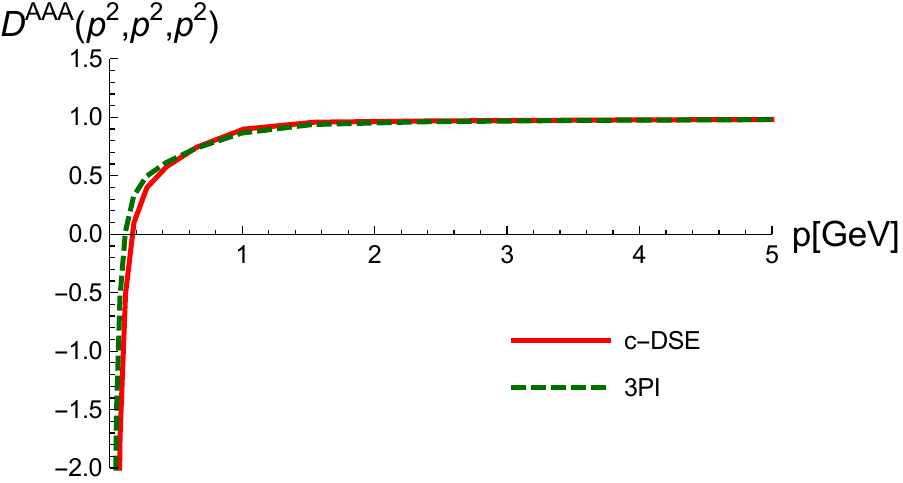}
 \caption{\textit{Left:} Gluon dressing function of three-dimensional Yang-Mills theory for the truncation shown in Fig.~\ref{fig:soe} and variations thereof. \textit{Right:} Three-gluon vertex dressing function from the system of DSEs (red, continuous line) and from the system of equations of motion of the 3PI effective action (green, dashed line). The results for a bare four-gluon vertex lie on top of the DSE results and are not shown.}
 \label{fig:3d_variations}
\end{figure}

\begin{figure}[tb]
 \includegraphics[width=0.49\textwidth]{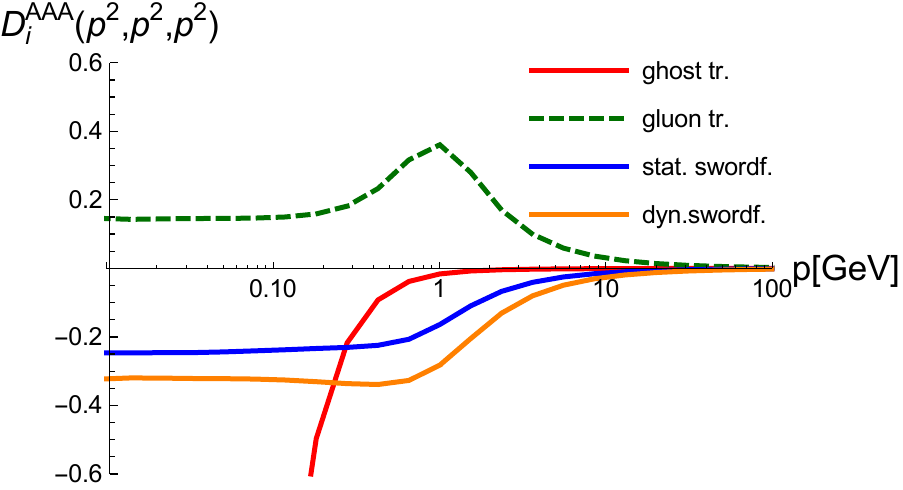}
 \hfill
 \includegraphics[width=0.49\textwidth]{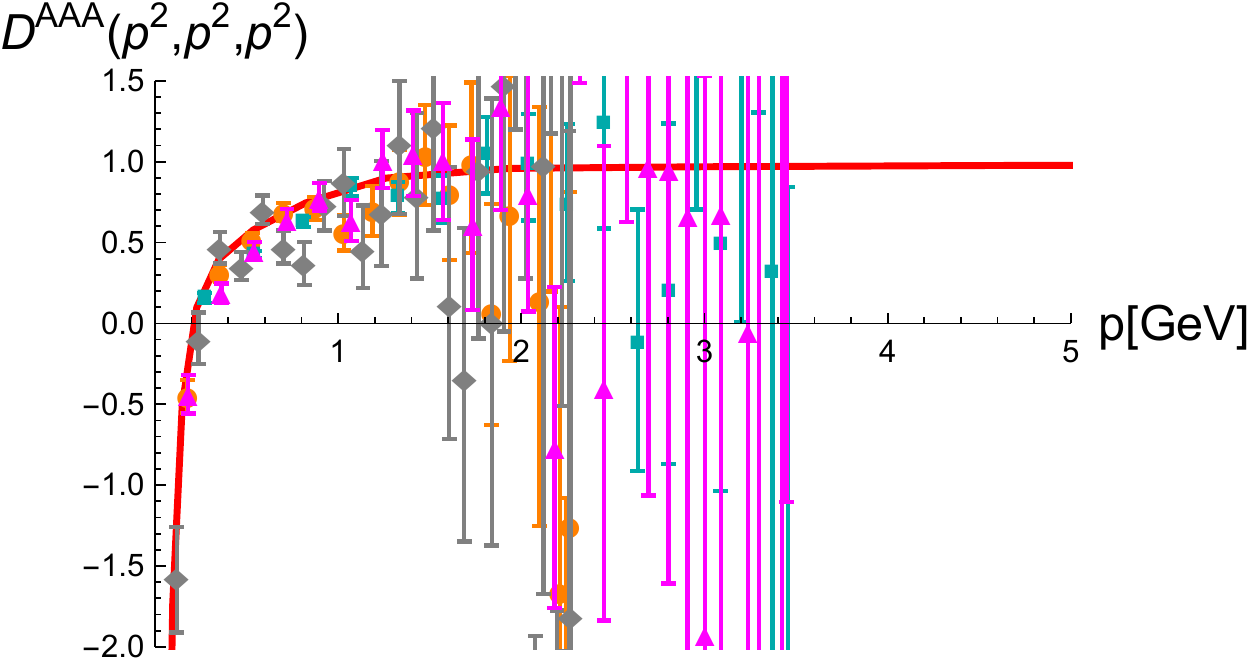}
 \caption{\textit{Left:} Contributions of single diagrams to the three-gluon vertex. \textit{Right:} Comparison of the result for the three-gluon vertex tree-level dressing function to lattice results \cite{Cucchieri:2008qm}.}
 \label{fig:3d_tg_cancellations}
\end{figure}

The most direct way to estimate the validity of a calculation with functional methods is at first sight a comparison with results from lattice simulations. Lattice sizes have become large enough to probe the IR regime a while ago \cite{Cucchieri:2007md,Cucchieri:2008fc,Sternbeck:2007ug,Bogolubsky:2009dc}, and the obtained results are often used to gauge the accuracy of results from other methods. However, such comparisons have to be taken with a grain of salt, since it is known that lattice results depend on the details of the employed gauge fixing algorithm \cite{Maas:2009se,Sternbeck:2012mf,Maas:2016frv}. Also for functional methods several solutions can be obtained \cite{Boucaud:2008ji,Fischer:2008uz,Alkofer:2008jy}, but the correspondence between the solutions of one method to the ones of the other are not known.
In addition, available lattice results are mostly limited to propagators in the vacuum and to the Landau gauge. However, gauges beyond the Landau gauge are attracting more interest lately, for example, the linear covariant gauges and the Coulomb gauge \cite{Huber:2014isa,Aguilar:2015nqa,Huber:2015ria,Bicudo:2015rma,Vastag:2015qjd,Campagnari:2016wlt,Aguilar:2016ock}.

In the light of this it becomes obvious that some other method of testing the reliability of results is required. One possibility is to probe modifications of a truncation, in particular to enlarge the truncation. Such tests are not as straightforward as they might seem. The reason is that when models are contained in a calculation, changes of a truncation might be compensated by readjusting their parameters. Thus, self-contained truncations are ideal where no model parameters appear. Here, such a truncation is investigated for three-dimensional Yang-Mills theory in the Landau gauge \cite{Huber:2016tvc,Huber:2016hns}. The basic truncation prescription is to retain only primitively divergent correlation functions. This truncation is complete at the perturbative one-loop level. The resulting equations for Dyson-Schwinger equations are depicted in Fig.~\ref{fig:soe}. In all calculations only the tree-level tensors are considered dynamically. For the ghost-gluon vertex this is sufficient for the Landau gauge. For the three-gluon vertex it was found in four dimensions that other dressing functions are small \cite{Eichmann:2014xya}. For the four-gluon vertex it was shown for some dressing functions that they are small when evaluating them a posteriori with the obtained solution as input \cite{Cyrol:2014kca}.

The reasons for working in three dimensions are related to the absence of renormalization. In four dimensions, renormalization introduces an anomalous running of the dressing functions. This leads to two problems in calculations with equations of motion as employed here. First, to reproduce the anomalous running in the gluon propagator, perturbative contributions beyond one-loop are required. Instead of including such contributions explicitly, they are typically included via an additional factor in the gluon loop \cite{vonSmekal:1997vx,Huber:2012kd}. However, this has some influence on the nonperturbative regime \cite{Huber:2014tva}. In three dimensions, on the other hand, there is no anomalous running and thus no resummation. Second, the employed cutoff regularization introduces spurious divergences which have a complicated dependence on the cutoff due to the anomalous running of the dressings \cite{Huber:2014tva}. Various methods to subtract them lead to quantitatively different results. Again, in three-dimensions this is different since the spurious terms are linear and logarithmic in the cutoff. They are subtracted here by fitting the corresponding coefficients with respect to the cutoff.

\begin{figure}[tb]
 \includegraphics[width=0.49\textwidth]{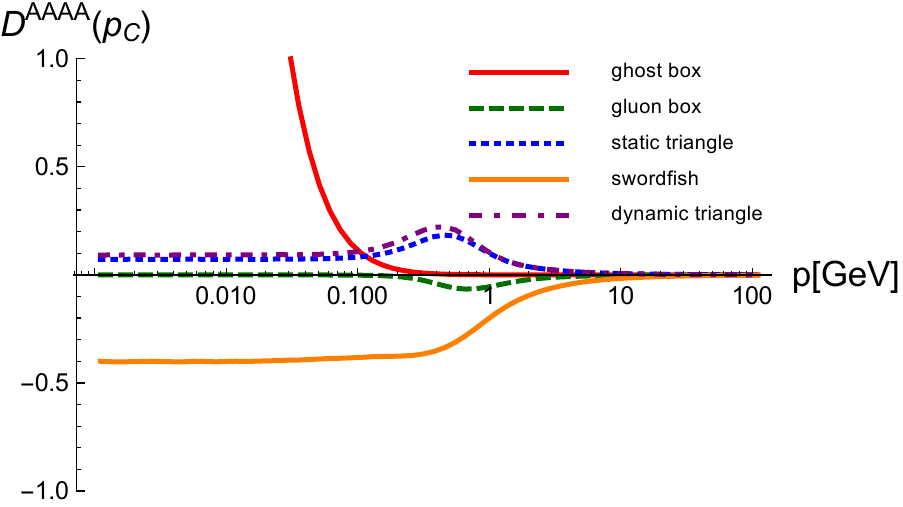}
 \hfill
 \includegraphics[width=0.49\textwidth]{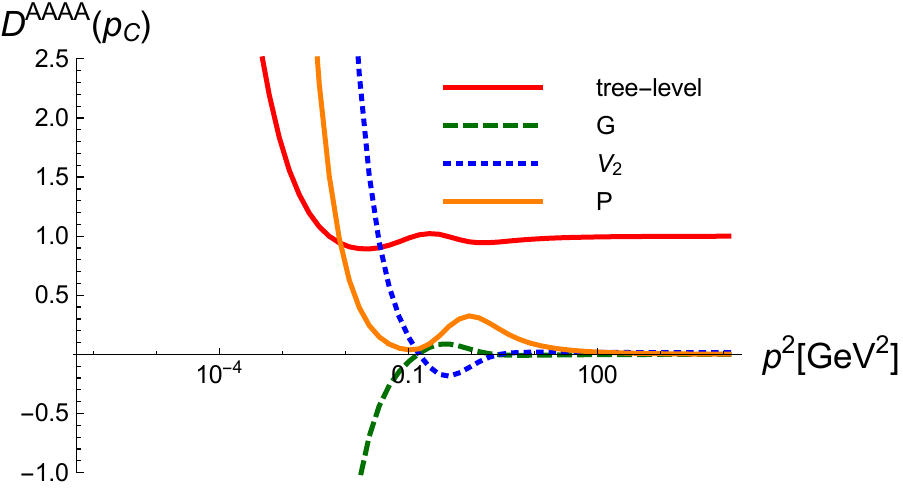}
 \caption{\textit{Left:} Contributions of single diagrams to the four-gluon vertex. \textit{Right:} Results for various dressings of the four-gluon vertex.}
 \label{fig:3d_fg_cancellations}
\end{figure}

\begin{figure}[tb]
 \centering \includegraphics[width=0.49\textwidth]{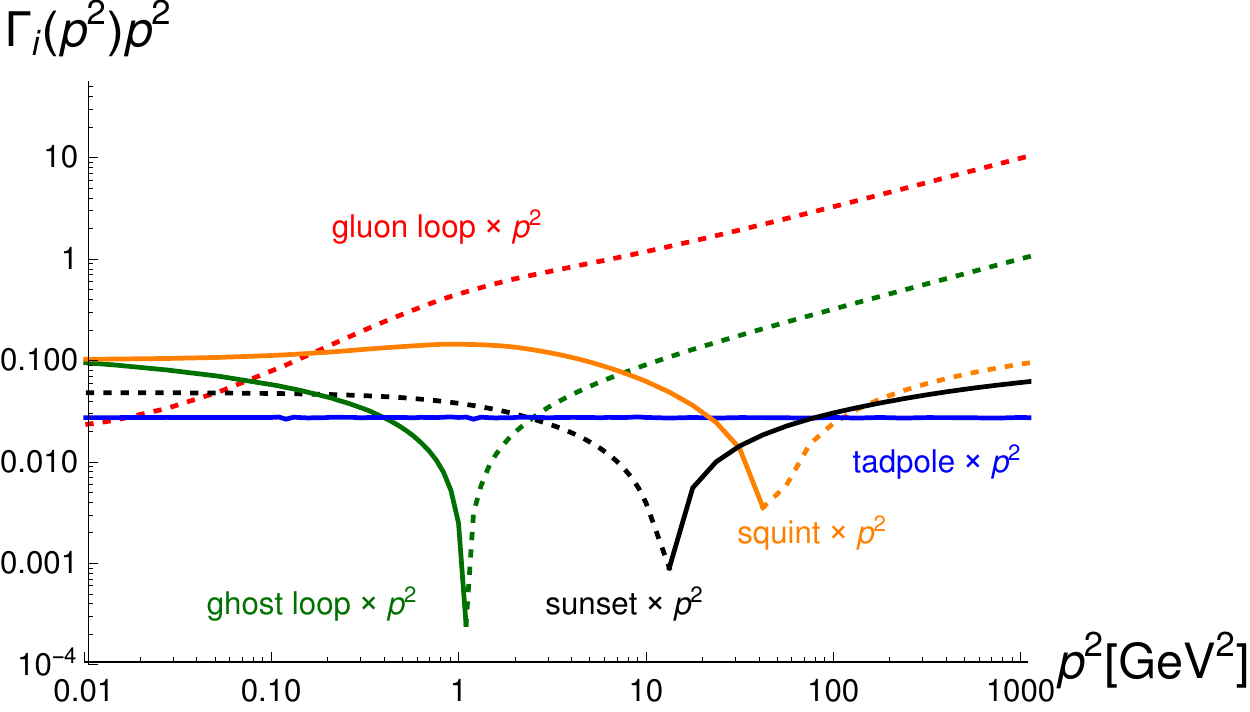}
 \hfill
 \centering \includegraphics[width=0.49\textwidth]{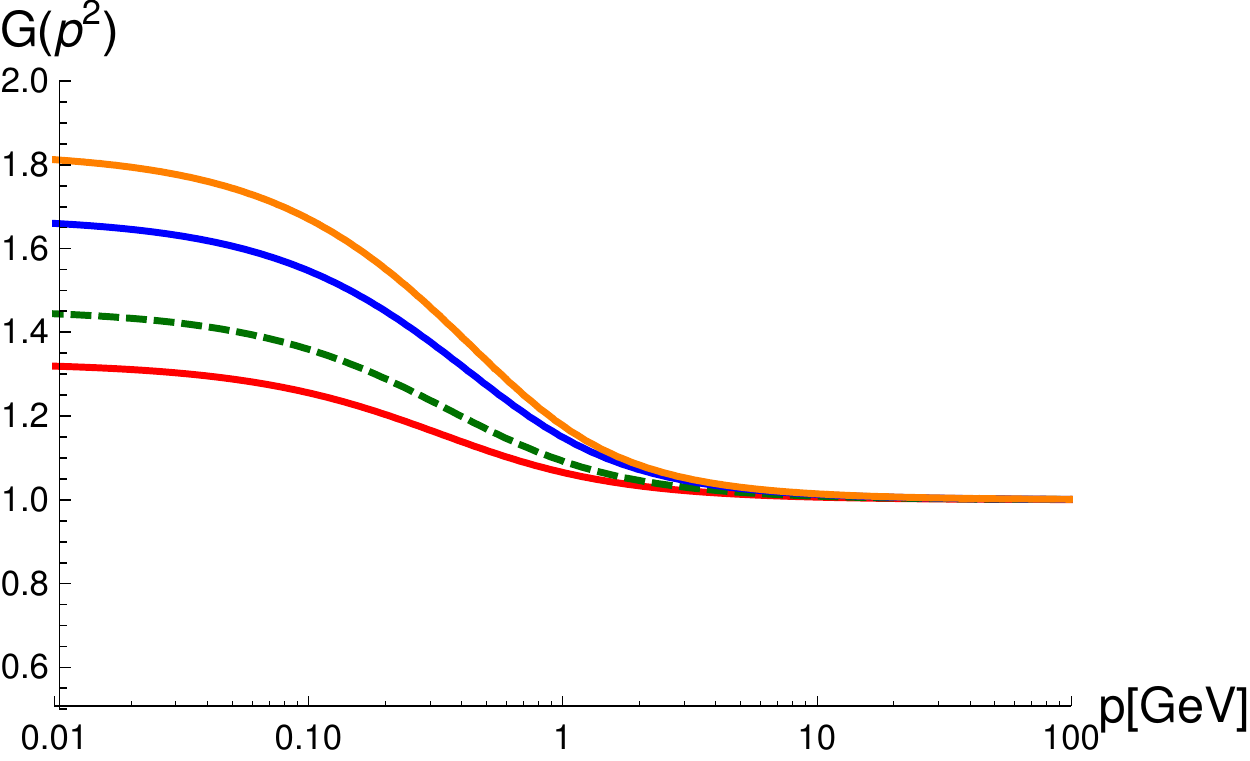}
 \caption{\textit{Left:} Contributions of single diagrams to the gluon dressing function. \textit{Right:} Family of solutions for the ghost dressing function obtained using different conditions for the gluon propagator in the IR.}
 \label{fig:3d_glSD_family}
\end{figure}

Results for the system of equations depicted in Fig.~\ref{fig:soe} are shown in Figs.~\ref{fig:3d_variations}, \ref{fig:3d_tg_cancellations}, and \ref{fig:3d_fg_cancellations}. One possibility to probe truncation effects is to employ the same truncation prescription for a different set of functional equations. In Fig.~\ref{fig:3d_variations}, the result for the gluon dressing function from its Dyson-Schwinger equation (DSE) is compared to the result from its equation of motion from the 3PI effective action. The observed difference in the gluon propagator is quite small, despite larger differences in the ghost-gluon vertex \cite{Huber:2016tvc}.
Another test is a change of the four-gluon vertex. Results from functional equations already showed in four dimensions \cite{Binosi:2014kka,Cyrol:2014kca} that the tree-level tensor follows the one-loop resummed perturbative behavior down to a few GeV and deviates considerably only at low momenta due to contributions from the ghosts. Furthermore, other dressing functions tested in \cite{Cyrol:2014kca} are very small. Similar results were found in three dimensions \cite{Huber:2016tvc}, see Fig.~\ref{fig:3d_fg_cancellations}. This raises the question what would happen if a bare vertex were used. As it turns out, the results do only change slightly for all other correlation functions, see, for example, the gluon dressing function in Fig.~\ref{fig:3d_variations}. With hindsight, the finding that the influence of the four-gluon vertex on the gluon propagator is not very large is not so surprising, since it enters in the sunset diagram. This diagram, however, is found to yield only a small contribution in general \cite{Huber:2016tvc}, see Fig.~\ref{fig:3d_glSD_family}. Indirectly, the four-gluon vertex could influence the gluon dressing function via the three-gluon vertex, the only other place where it enters in this truncation. However, it turns out that the three-gluon vertex changes almost not at all \cite{Huber:2016tvc}.

An interesting observation is that the three- and the four-gluon vertices are close to the tree-level and only deviate from that at very low momenta where they diverge linearly. This behavior sets in at so low momenta that it has hardly any influence on other dressing functions. As it turns out, the reason for the small deviations from the tree-level is cancellations between the diagrams, see Figs.~\ref{fig:3d_tg_cancellations} and \ref{fig:3d_fg_cancellations}. If this behavior persists for higher correlation functions, they will naturally be very small since they are zero at tree-level.

The question of the realization of a family of solutions is a little bit more tricky than in four dimensions due to the absence of renormalization. In four dimensions, a solution can be chosen by the renormalization condition for the ghost propagator \cite{Boucaud:2008ji,Fischer:2008uz,Alkofer:2008jy}. Motivated by results from the functional renormalization group in four dimensions, where spurious divergences can be handled via tuning of a bare mass parameter \cite{Cyrol:2016tym}, one can try and fix different values for the gluon propagator at zero momentum as an alternative way to handle spurious divergences. Calculating only the propagators with bare vertices as input, it is indeed found that this leads to the expected behavior of the ghost dressing function \cite{Huber:2016hns}, see Fig.~\ref{fig:3d_glSD_family}.

\section{Non-vanishing temperature}
\label{sec:non-zero_temp}

\begin{figure}[tb]
 \includegraphics[width=0.49\textwidth]{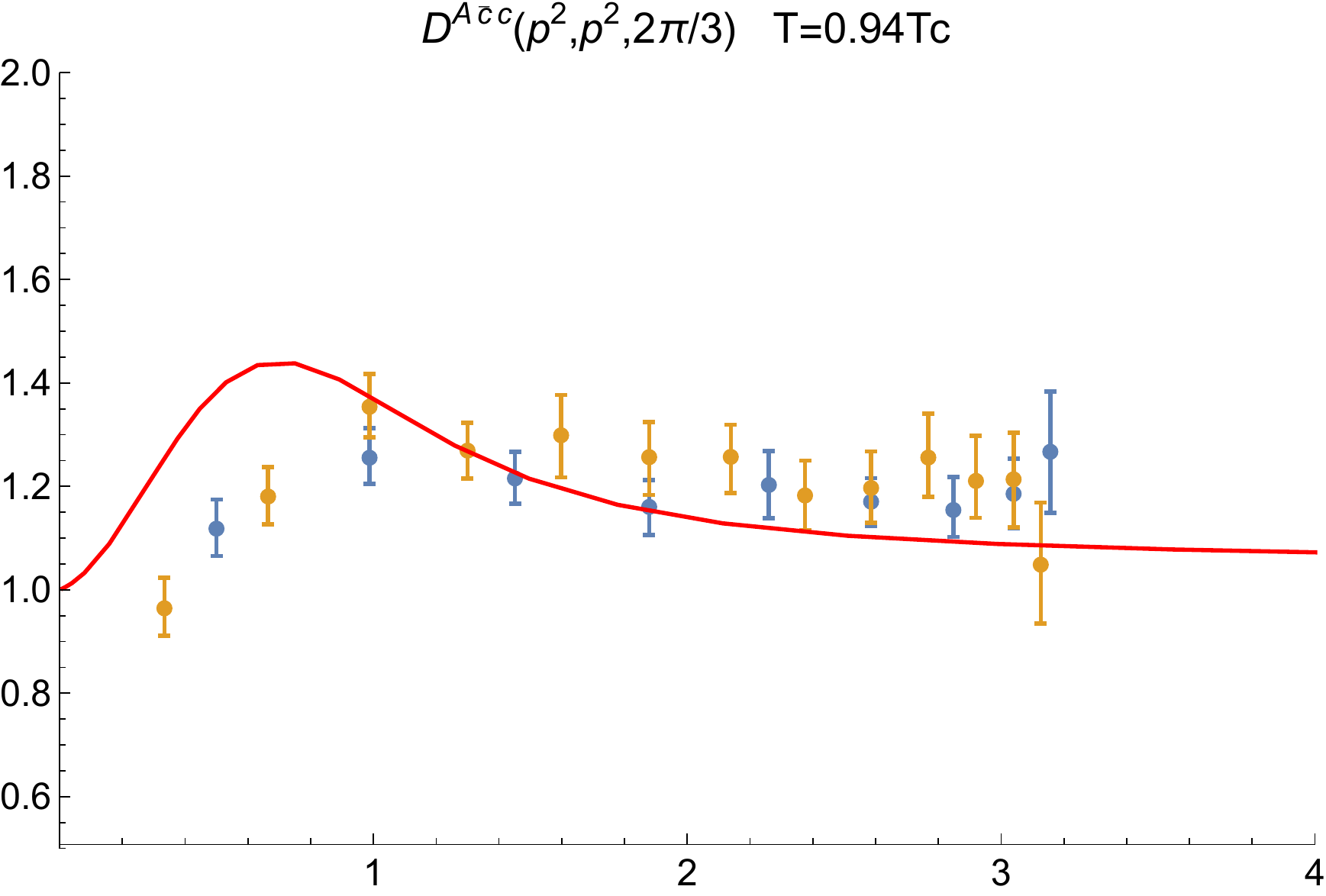}
 \hfill
 \includegraphics[width=0.49\textwidth]{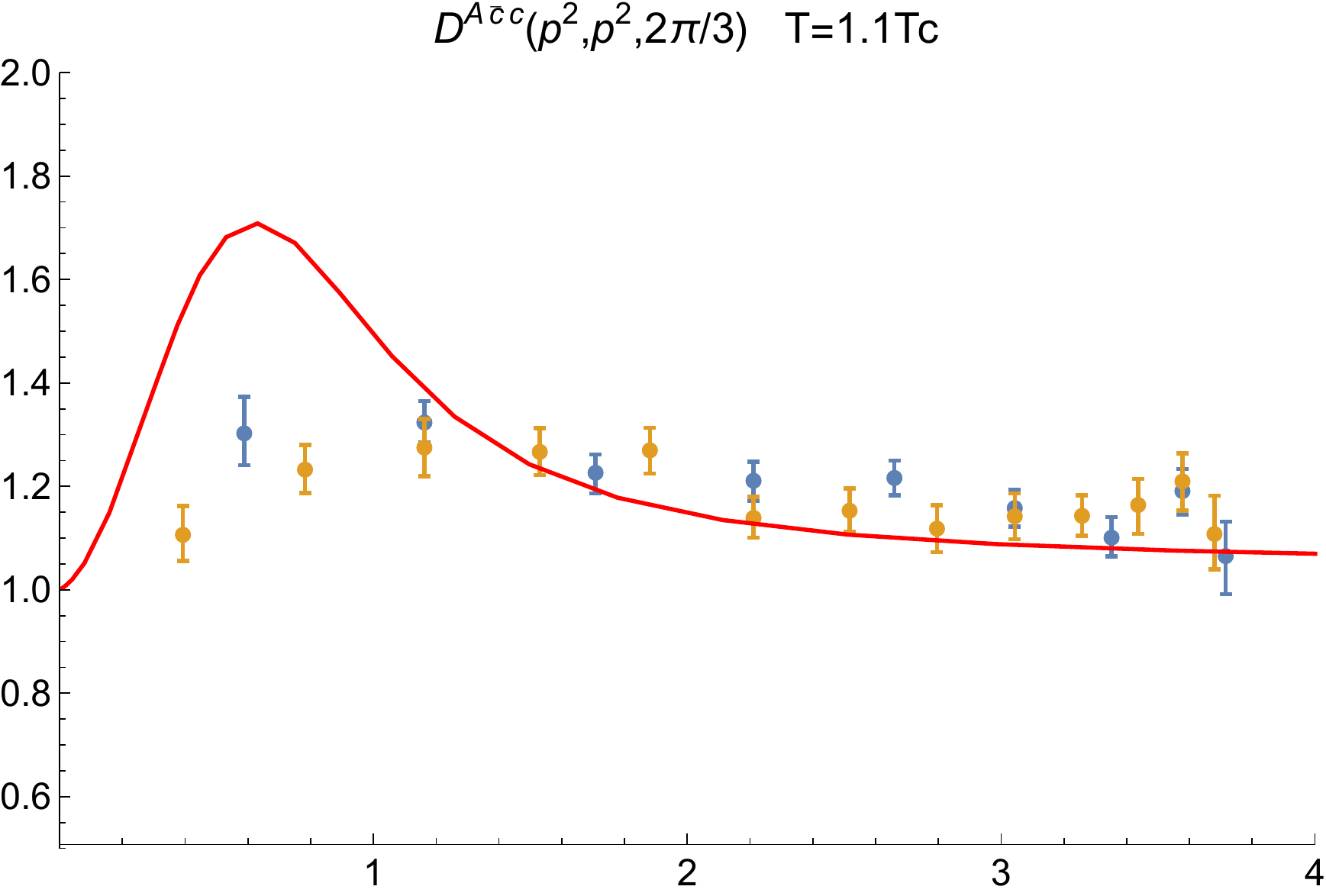}
 \caption{Ghost-gluon vertex dressing function at the symmetric point in comparison to lattice results \cite{Fister:2014bpa}.}
 \label{fig:ghgL}
\end{figure}

\begin{figure}[tb]
 \includegraphics[width=0.49\textwidth]{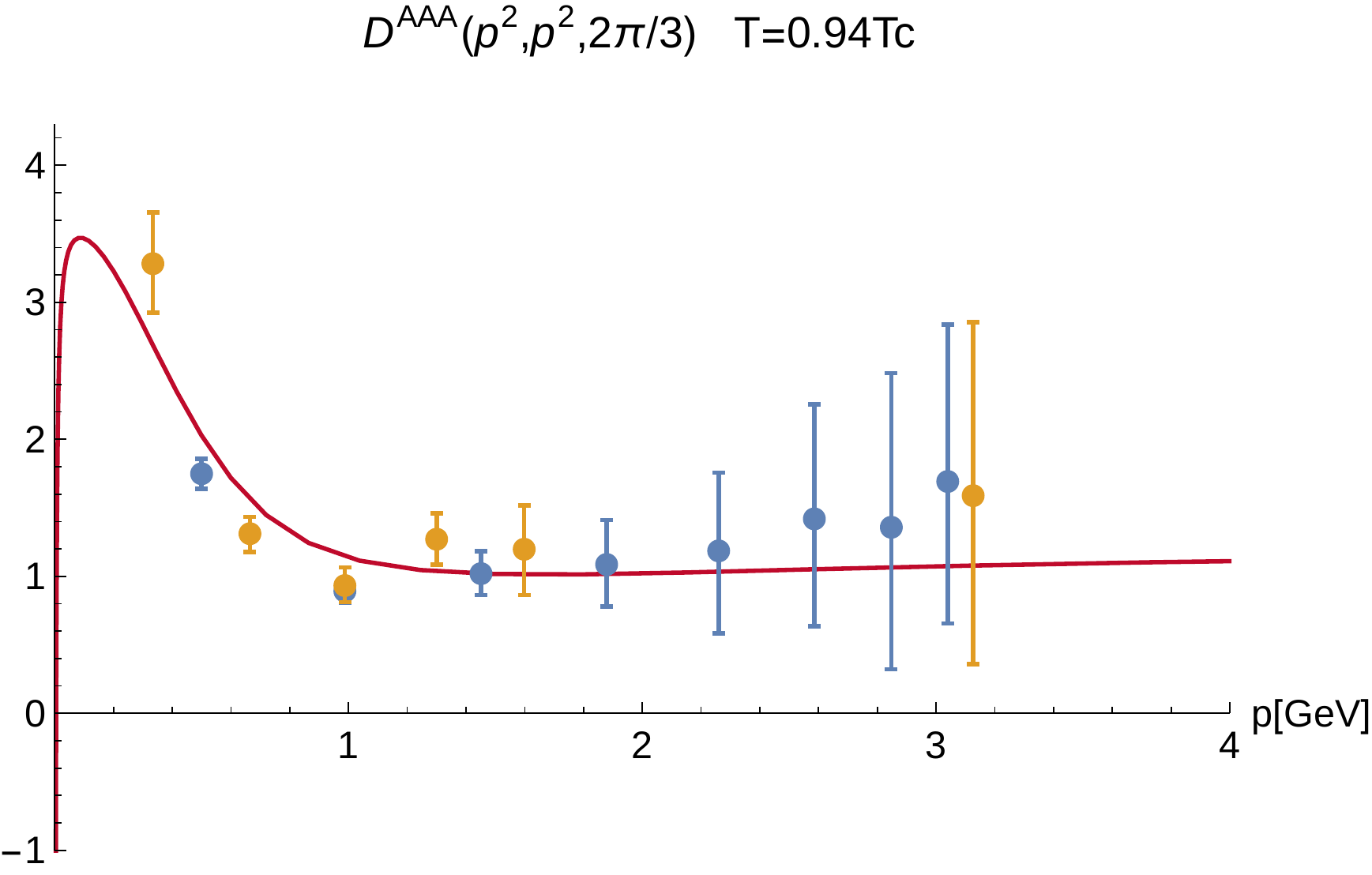}
 \hfill
 \includegraphics[width=0.49\textwidth]{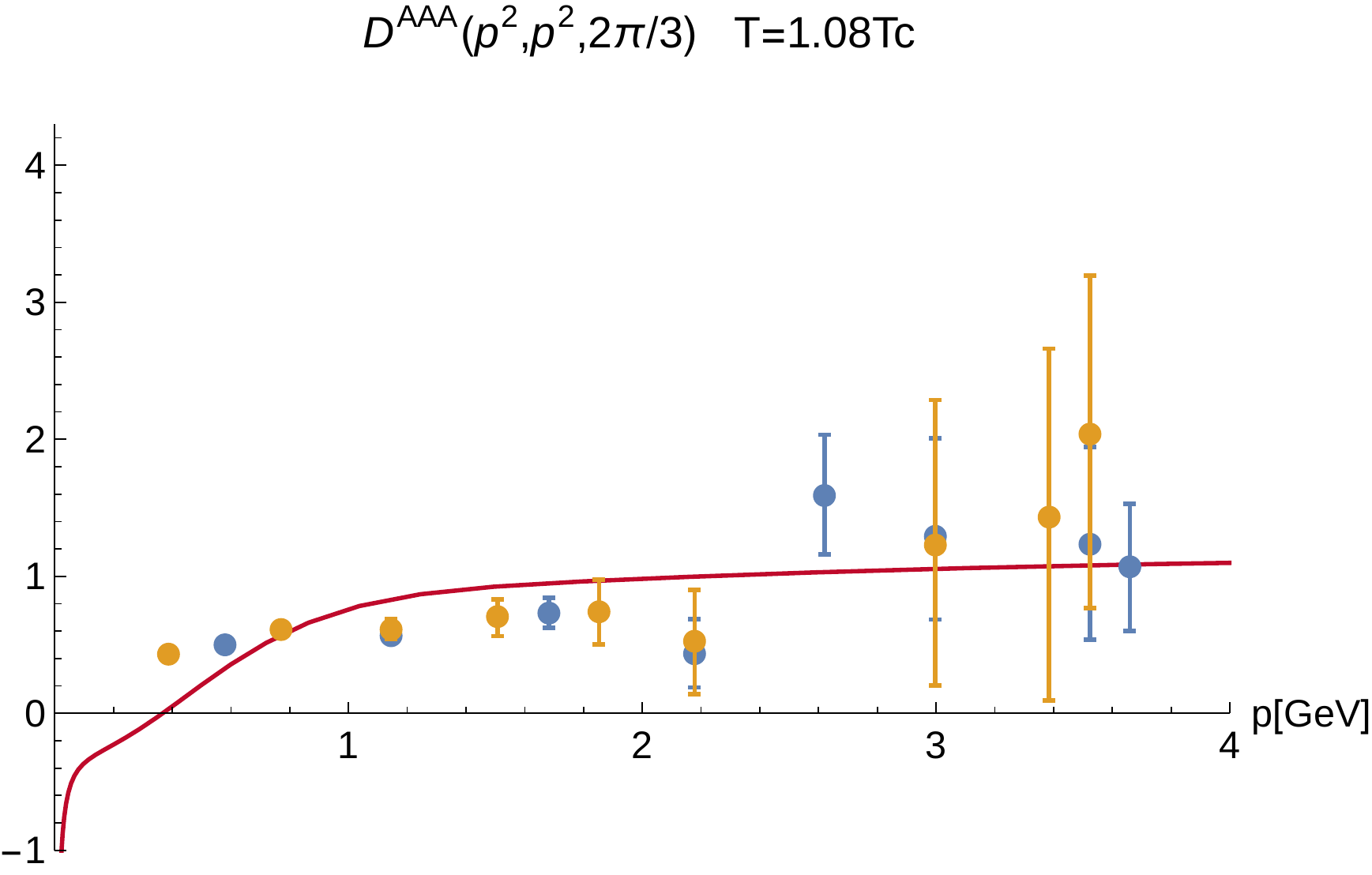}
 \caption{Three-gluon vertex dressing function at the symmetric point in comparison to lattice results \cite{Fister:2014bpa}.}
 \label{fig:tgL}
\end{figure}

At non-vanishing temperature, the gluon propagator splits into chromoelectric and chromomagnetic parts. The number of contributions in functional equations containing gluons multiplies accordingly. As at zero temperature, the gluon propagator is plagued by spurious divergences. They can be dealt with, for example, by using explicit additional renormalization conditions \cite{Quandt:2015aaa}. Calculations of the propagators can be found, e.g., in \cite{Maas:2005hs,Cucchieri:2007ta,Fister:2011uw,Quandt:2015aaa}. Here I aim at the calculation of three-point functions. As input, I use fits of lattice data \cite{Fischer:2010fx,Maas:2011ez} for the gluon propagator and results for the ghost propagator calculated from the gluon propagator fits with a bare ghost-gluon vertex \cite{Huber:2013yqa}. The ghost propagator results show reasonable agreement with lattice results \cite{Huber:2013yqa}.

\begin{figure}[tb]
 \includegraphics[width=0.49\textwidth]{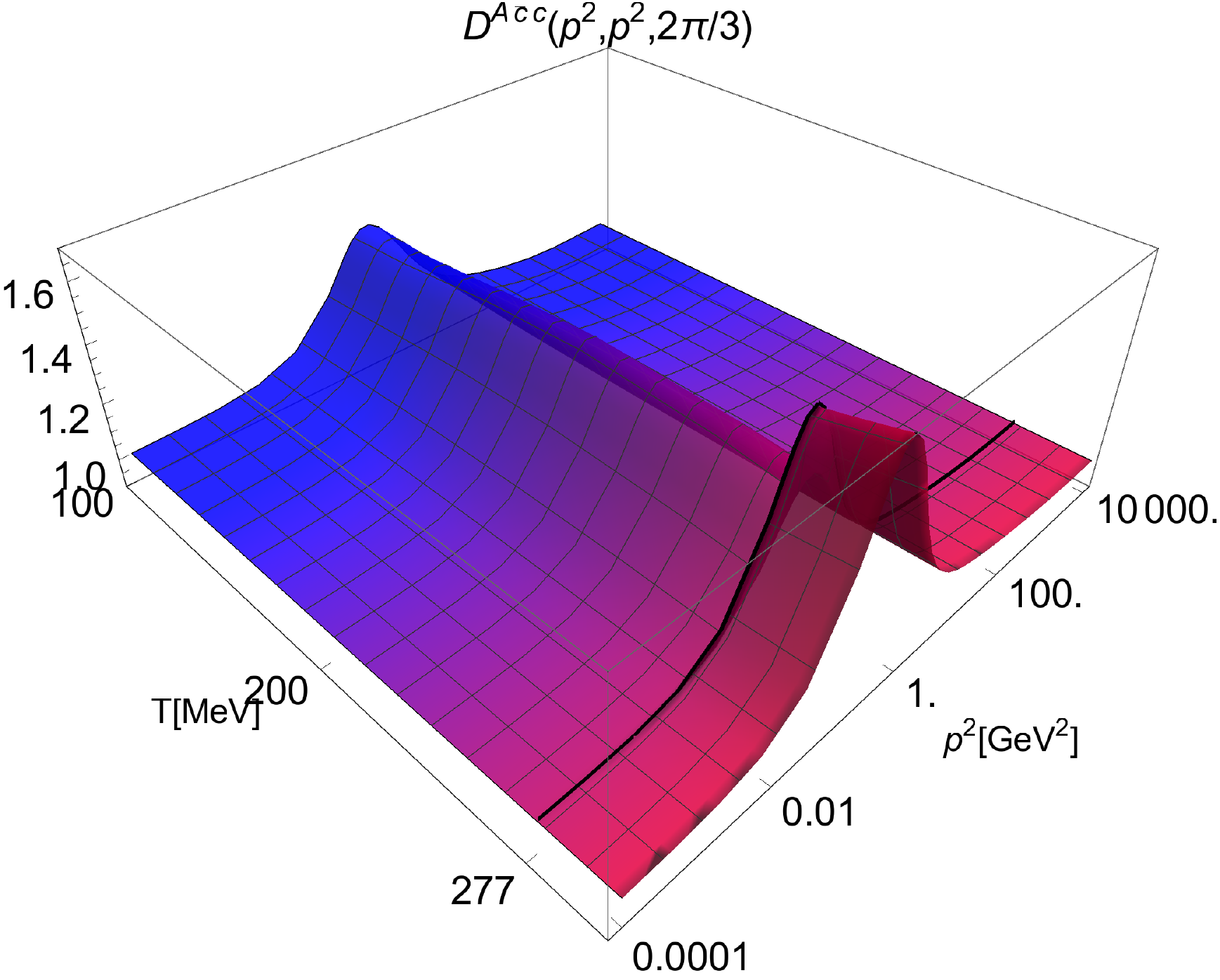}
 \hfill
 \includegraphics[width=0.49\textwidth]{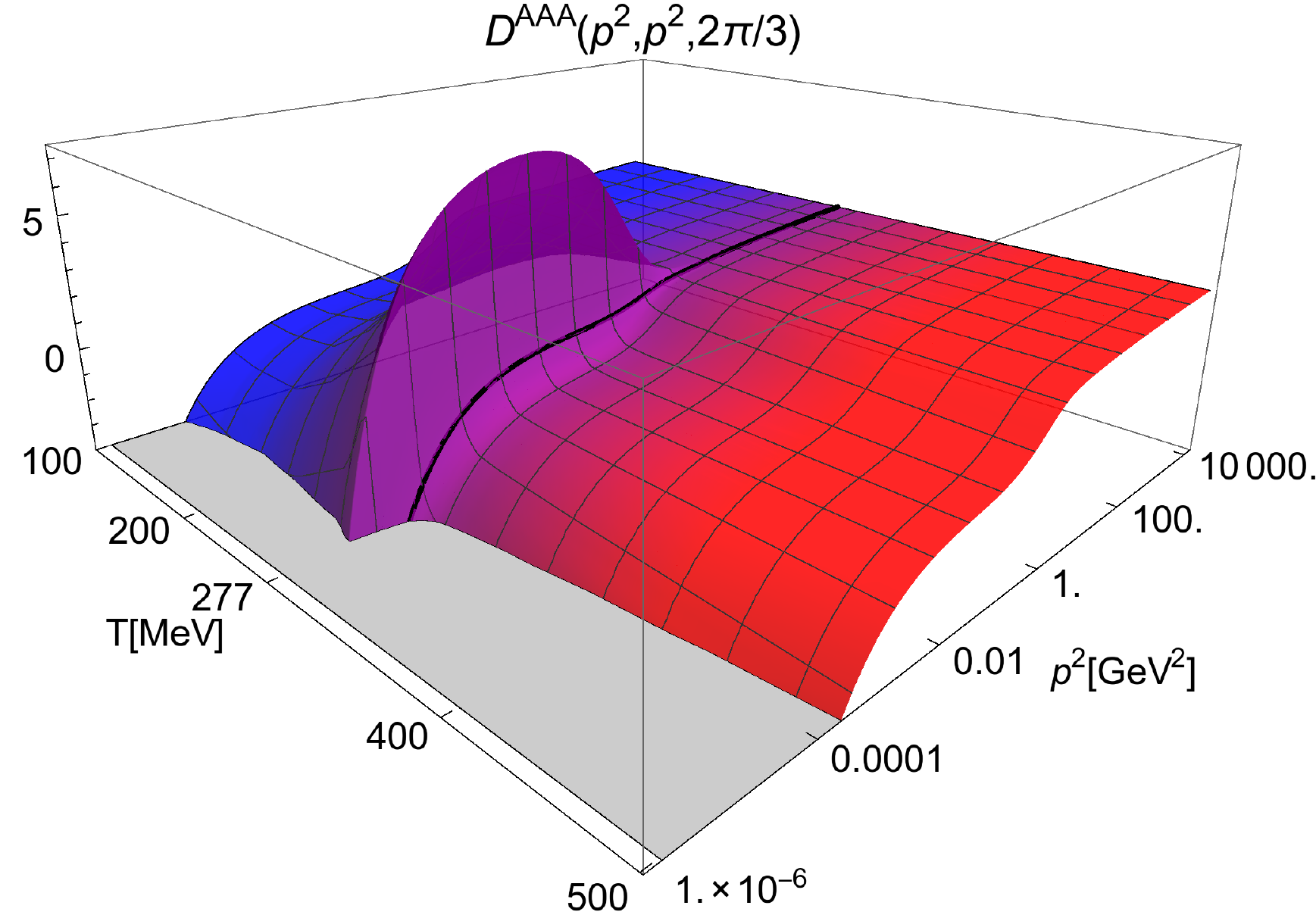}
 \caption{\textit{Left:} Temperature dependence of the ghost-gluon vertex dressing function at the symmetric point.
 \textit{Right:} Temperature dependence of the three-gluon vertex dressing function at the symmetric point. In both plots the black line denotes the phase transition.}
 \label{fig:3p_3d}
\end{figure}

The ghost-gluon vertex was calculated with the approximation of only considering the zeroth Matsubara frequencies so that solely the chromomagnetic dressing function remains. The resulting DSE was solved self-consistently for the gauge group $SU(3)$. The results for two temperatures are compared to $SU(2)$ lattice results in Fig.~\ref{fig:ghgL}. The agreement is in general not very good, but similar discrepancies are known for $SU(2)$ in the vacuum \cite{Huber:2012kd,Huber:2016tvc}. The full temperature dependence for one momentum configuration is shown in Fig.~\ref{fig:3p_3d}. In contrast to the results shown in \cite{Huber:2013yqa}, the temperature dependence is smooth. The reason is that the fit parameters for the gluon dressing function show some fluctuations. Instead of using the parameters directly as done in \cite{Huber:2013yqa}, here the parameters are fitted to guarantee a smooth behavior \cite{Luecker:2013th}.

The three-gluon vertex was approximated analogously to the ghost-gluon vertex and only the projection with the chromomagnetic tree-level tensor was calculated. Furthermore, the shown results correspond to a semi-perturbative approximation, viz. the available input for the propagators was used, but the equation was not solved self-consistently. In Fig.~\ref{fig:tgL} the results are compared to lattice results. From lattice simulations it was known that the vertex dressing shows an enhancement in the nonperturbative regime below the phase transition \cite{Fister:2014bpa}. However, as there is strong evidence that at zero temperature and in the infinite temperature limit the vertex becomes negative \cite{Mendes:2008ux,Huber:2012kd,Pelaez:2013cpa,Aguilar:2013vaa,Blum:2014gna,Athenodorou:2016oyh,Duarte:2016ieu,Sternbeck:2016tgv}, it is not clear from the lattice data if this changes for intermediate temperatures or if the vertex goes down again at even lower momenta. The calculation of its DSE favors the second possibility as a clear zero crossing is observed, although at very low momenta, see Fig.~\ref{fig:3p_3d}. In addition, already this simple setup for the vertex shows even quantitative agreement with lattice results, see Fig.~\ref{fig:tgL}. To find out if this is a mere coincidence will require more sophisticated calculations. A useful finding of this calculation concerns the importance of various contributions. For example, the gluon triangle can be considered as twelve single contributions distinguished by which gluon propagators appear. It was found that the contributions with mixed propagators are negligible compared to the others. Furthermore, above the phase transition the contribution with only chromoelectric propagators becomes very small. If this observation holds for a fully dynamical three-gluon vertex, dropping all mixed contributions would be an extremely good approximation but reduce the computing time considerably.

\section{Conclusions}

In Sect.~\ref{sec:truncations} possibilities to test truncations of functional equations have been discussed and corresponding results for three-dimensional Yang-Mills theory have been shown. The main finding was that a self-contained truncation including all primitively divergent correlation functions performs well under such tests. It was found that gluonic vertices stay close to the tree-level with the main deviations stemming from the ghost contributions in the deep IR. The reason for this is a cancellation mechanism between the other gluonic contributions.  A study of dressing functions beyond the tree-level dressing calculated here is one possibility for future extensions, as are calculations of neglected, not primitively divergent four-point functions.

Results for the ghost-gluon and three-gluon vertices at non-zero temperature have been presented in Sec.~\ref{sec:non-zero_temp}. As input, fits for the gluon dressing functions to lattice results and the ghost dressing function calculated from these fits were used. While the employed setup for the three-gluon vertex is still rather simple, it already reproduces the main features of the vertex at non-zero temperature seen previously in lattice calculations. These features are an enhancement below the phase transition and the existence of a zero crossing for all temperatures. The results for the ghost-gluon vertex, on the other hand, do not show a large temperature dependence of this quantity.

\section{Acknowledgments}

Computation were performed with \textit{DoFun} \cite{Huber:2011qr,Alkofer:2008nt} and \textit{CrasyDSE} \cite{Huber:2011xc} using HPC Clusters at the University of Graz. Feynman diagrams were created with Jaxodraw \cite{Binosi:2003yf}.
Support by the FWF (Austrian science fund) under Contract No. P27380-N27 is gratefully acknowledged.

\bibliography{literature_Conf12}

\end{document}